\documentclass[preprint]{aastex}
\doublespace
\usepackage{graphicx}
\shortauthors{Pal'shin et al.}

\begin{document}

\title{Konus-\textit{Wind} and Helicon-\textit{Coronas-F} Observations of Solar Flares}

\author{V. D. Pal'shin\altaffilmark{1,2}, Yu. E. Charikov\altaffilmark{1,2,3}, R. L. Aptekar\altaffilmark{1}, %
S. V. Golenetskii\altaffilmark{1}, A. A. Kokomov\altaffilmark{1}, \\%
D. S. Svinkin\altaffilmark{1}, Z. Ya. Sokolova\altaffilmark{1}, %
M. V. Ulanov\altaffilmark{1,2}, D. D. Frederiks\altaffilmark{1}, and
A. E. Tsvetkova\altaffilmark{1}}
\altaffiltext{1}{Ioffe Physical Technical Institute, St. Petersburg,
Russian Federation \email{val@mail.ioffe.ru}}
\altaffiltext{2}{St. Petersburg State Polytechnical University, St.
Petersburg, Russian Federation}
\altaffiltext{3}{Central (Pulkovo) Astronomical Observatory, Russian
Academy of Sciences, St. Petersburg, Russian Federation}

\begin{abstract}
Results of solar flare observations obtained in the
Konus-\textit{Wind} experiment from November, 1994 to December, 2013
and in the Helicon-\textit{Coronas-F} experiment during its
operation from 2001 to 2005, are presented. For the periods
indicated Konus-\textit{Wind} detected in the trigger mode 834 solar
flares, and Helicon-\textit{Coronas-F} detected more than 300 solar
flares.

A description of the instruments and data processing techniques are
given. As an example, the analysis of the spectral evolution of the
flares SOL2012-11-08T02:19 (M1.7) and SOL2002-03-10T01:34 (C5.1) is
made with the Konus-\textit{Wind} data and the flare
SOL2003-10-26T06:11 (X1.2) is analyzed in the 2.223 MeV deuterium
line with the Helicon-\textit{Coronas-F} data.

\end{abstract}



\section{INTRODUCTION}
Studying the dynamics of the energy spectra of X-ray and gamma-ray
emission of solar flares is a powerful tool for the investigation of
charged particle acceleration and the diagnostics of flare magnetic
loop plasma (see review of \citet{Fletcher2011} and references
therein). Since 2002, the main results from observations of solar
flares in the hard X-ray and gamma-ray ranges have been obtained
with the \textit{RHESSI} spectrometer \citep{Lin2002}, which has
excellent energy and angular resolution. Operating since 2008
onboard the \textit{Fermi} observatory the Gamma-Ray Burst Monitor
(GBM) \citep{Meegan2009} is also carrying out observations of solar
flares in a range from $\sim$8 keV to $\sim$40 MeV. The Russian
gamma-ray spectrometer Konus, which successfully operates onboard
the U.S. \textit{Wind} spacecraft, has a number of advantages when
compared with \textit{RHESSI} and GBM, operating on near-Earth
spacecraft. The Konus-\textit{Wind} experiment has been continuously
carried out since November 1994 in the favorable conditions of
interplanetary space far from the Earth's magnetosphere, which
ensures no interference from Earth's radiation belts nor occulting
by the Earth. Owing to such an orbit a useful observation time is
$\sim$95\%, while the configuration of detectors provides continuous
monitoring of the entire celestial sphere in a wide energy range
from $\sim$18 keV to $\sim$15 MeV under extremely stable background
(in the absence of powerful streams of charged particles from the
Sun). In addition, the effective area of the spectrometer in the
range $>$100 keV may be more than the effective area of the
\textit{RHESSI} detector (for certain configurations) and resolution
in the MeV range is better than that of the GBM detectors. During
the operation of Konus-\textit{Wind} and other series of the Konus
experiments a unique database on solar flares has been accumulated.
The aim of this paper is to describe the obtained data, the methods
of their processing, as well as to present the results of data
analysis through examples on specific flares.

\section{INSTRUMENTS}
\subsection{Konus-\textit{Wind}}
\label{secKW}
The Russian-American experiment Konus-\textit{Wind} is designed to
study the temporal and spectral characteristics of gamma-ray bursts,
soft gamma repeaters and other transient phenomena in a wide energy
range from $\sim$18 keV to $\sim$15 MeV. The instrument consists of
two spectrometric gamma-ray detectors NaI(Tl) (13 cm in diameter,
7.5 cm in height). Their axes are directed to the south and the
north ecliptic pole, respectively, that provides a continuous
monitoring of the entire celestial sphere. Each detector has an
effective area of $\simeq 80-160$~cm$^2$, depending on the energy
and incident angle of a gamma quantum. Resolution of the
spectrometer is about 8\% (FWHM) at the 662 keV line, the
sensitivity of $\sim(1-5) \times 10^{-7}$~erg~cm$^{-2}$. Detectors
operate in two modes: ``Background'' and ``Burst'' (trigger mode).
In the ``Background'' mode a counting rate is measured in three
energy channels G1, G2, G3 with the nominal ranges 10--50, 50--200,
and 200--750~keV ($\sim$18--70, 70--300, 300--1160~keV in 2013
year), with a temporal resolution of 2.944~s. In the ``Burst'' mode
time history is measured in the same three channels with a
resolution varying from 2 to 256~ms and a total record duration of
230~s, and also 64 spectra are measured in two partially overlapping
bands with nominal boundaries 10--750~keV (PHA1) and 0.2--10 MeV
(PHA2) ($\sim$18--1160 keV and $\sim$0.3--15~MeV in 2013). Each band
has 63 channels. The first four spectra are measured with a fixed
accumulation time of 64~ms. For the next 52 spectra a system,
adapting spectrum measurement duration to the current emission
intensity, determines accumulation times, which may vary from 0.256
to 8.192~s. The last 8 spectra are measured with an accumulation
time of 8.192~s. As a result, the total duration of spectral
measurements can vary from 79.104~s for the most intense events up
to 491.776~s for typical bursts. Transition into the trigger mode
occurs at a statistically significant excess above a background
count rate within an interval of 1~s or 140~ms in the G2 energy
channel. Thus, a background count rate is determined at the
preceding interval of 30~s length. At the end of a trigger record
information is slowly rewritten from the instrument's memory to the
on-board memory, which takes 1--1.5~hours. While rewriting the
operation of the instrument in the ``Background'' mode stops, no new
trigger can be generated at this time either, however one of the
backup systems continues to transmit G2 intensity measurements
through a ``housekeeping'' telemetry channel with a resolution of
3.680~s. A detailed description of the experiment is given in
\citet{Aptekar1995}, and the main results of gamma-ray burst and
soft gamma-ray repeater studies are described in a review by
\citet{Mazets2012}.

\subsection{Helicon-\textit{Coronas-F}}
\label{SecHel}
The Helicon gamma-ray spectrometer was one of the instruments
onboard the \textit{Coronas-F} solar space observatory
\citep{Oraevskii2002}, which had been in operation from August 2001
to December 2005 in a near-Earth low-eccentricity polar orbit
(inclination $82\fdg5$, distance from the Earth 500--550 km). The
spacecraft was stabilized by rotation with respect to the axis
directed toward the Sun within 10\arcmin. The spectrometer consisted
of two detectors, similar to those of Konus-\textit{Wind}, one of
which was oriented toward the Sun and the other viewed the antisolar
hemisphere. The ``Burst'' mode was similar to the
Konus-\textit{Wind} trigger mode; in the ``Background'' mode a time
history was measured in 8 energy channels covering the 10--200~keV
range with a time resolution of 1~s and 256-channel spectra were
measured in the 0.2--10~MeV range with accumulation time of 33.6~s.
Data output to the onboard memory was made without interruption in
measurements.

\section{OBSERVATIONS OF SOLAR FLARES IN KONUS-WIND EXPERIMENT }
\begin{figure}
\centering
\includegraphics[width=0.8\textwidth]{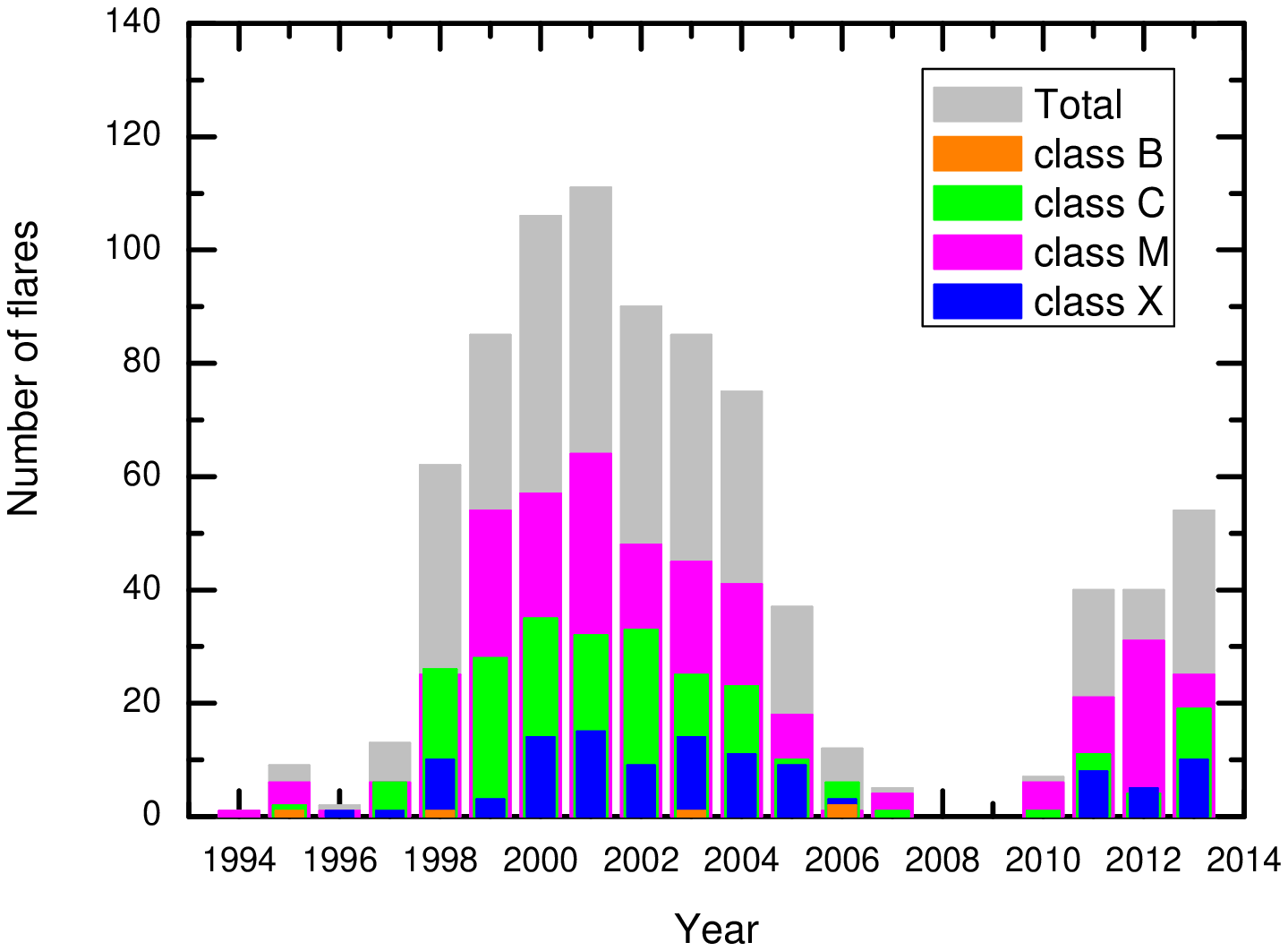}
\caption{Distribution of solar flares detected by
Konus-\textit{Wind} in the trigger mode over time.}
\label{Fig1}
\end{figure}

During the period from its launch in November 1994 until the end of
2013 Konus-\textit{Wind} had registered in the ``Burst'' mode 834
solar flares with the GOES classification: 113 X-class, 454 M-class,
262 C-class and 5 B-class. Figure~\ref{Fig1} shows the distribution
of the number of flares over the years. It is seen that the largest
number of flares occurred in years of maximum solar activity
2000--2002, while in 2008 and 2009 corresponding to the minimum of
solar activity, there were no sufficiently intense flares to cause a
trigger.

Due to the nature of the trigger algorithm (see
Section~\ref{secKW}), the transition to the ``Burst'' mode takes
place at appearing in a flare of high-energy radiation with quite a
rapid increase in intensity, wherein a smoother and softer initial
rise phase is skipped (and registered only in the background
record). Besides, the duration of many intense flares exceeds the
duration of measurements in the trigger mode. As a result, for a
significant proportion of long and intense flares, spectral data and
time profiles of high resolution are available only for part of a
flare.

\subsection{Analysis of Konus-\textit{Wind} solar flares}
Spectral analysis of the Konus-\textit{Wind} data is performed using
the software package XSPEC \citep{Arnaud1996}, for this purpose
spectra and a detector response matrix are converted into
fits-format. The incident angle on the detector is 90$^\circ$ for
solar flares, this entails the strong absorption of soft radiation
when passing through the aluminum container 2 mm thick housing the
scintillator. Despite this, the intensity of radiation from powerful
flares in the soft range may cause the overflow of the G1 channel
counter and significant distortion of spectra.

In the standard analysis of Konus-\textit{Wind} data a spectrum,
averaged over an interval of $\sim$100--300~s after the event, is
selected as the background spectrum, but the radiation from solar
flares often lasts longer than the measurement time of spectra in
the ``Burst'' mode (maximum 472~s), i.e. even last spectra of a
trigger record in such cases cannot be taken for background. Then,
it is often possible to use some spectra of neighboring trigger
events as the background if the background has not changed
significantly over the time interval between the analyzed flare and
neighboring trigger (it can be controlled by the level of background
in three channels: G1, G2, G3).

The XSPEC package contains a variety of standard spectral models,
but there are no standard models of thin and thick target, often
used for the analysis of spectra of solar flares. These models have
been added by us to XSPEC based on analogous models from the OSPEX
package, designed for the analysis of solar data.

\begin{figure}
\centering
\includegraphics[width=0.8\textwidth]{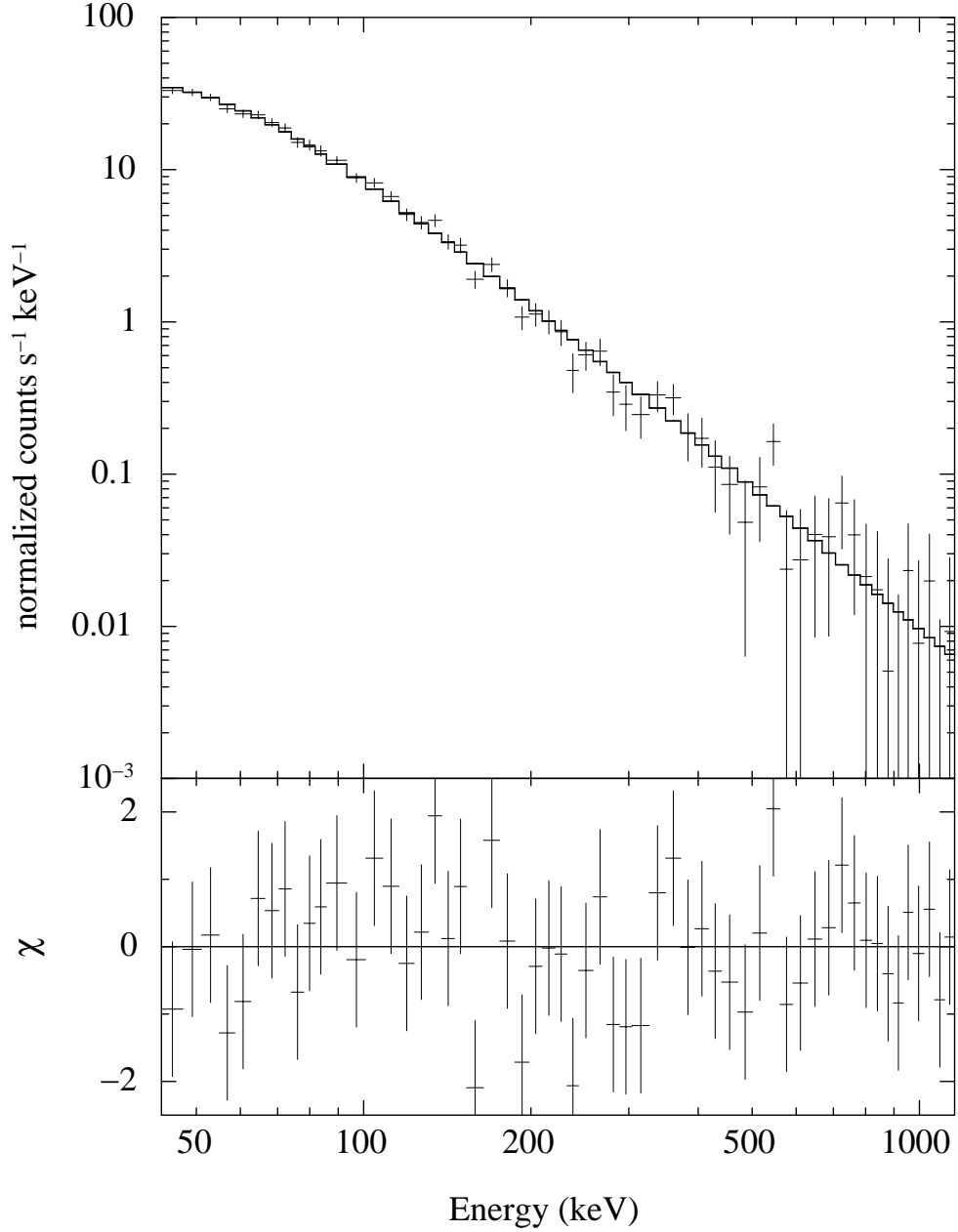}
\caption{Spectrum of the flare SOL2002-03-10T01:34 (see
section~\ref{SecSOL20020310}, measured by Konus-\textit{Wind} during
the interval 0--5.632~s. The spectrum is well described by a
power-law model with the index $\gamma = 2.65 \pm 0.30$
($\chi^2$=46/55 dof); significant emission is seen up to
$\sim$1~MeV.}
\label{Fig2}
\end{figure}

In the $>$40~keV range, where the contribution of the thermal
component of radiation is usually small, flare spectra are often
well described by a simple power-law model: $dN/dE \propto
E^{-\gamma}$. An example of such a spectrum measured by
Konus-\textit{Wind} is shown in Fig.~\ref{Fig2}.

\subsection{Flare SOL2012-11-08T02:19}
\begin{figure}
\centering
\includegraphics[width=0.47\textwidth]{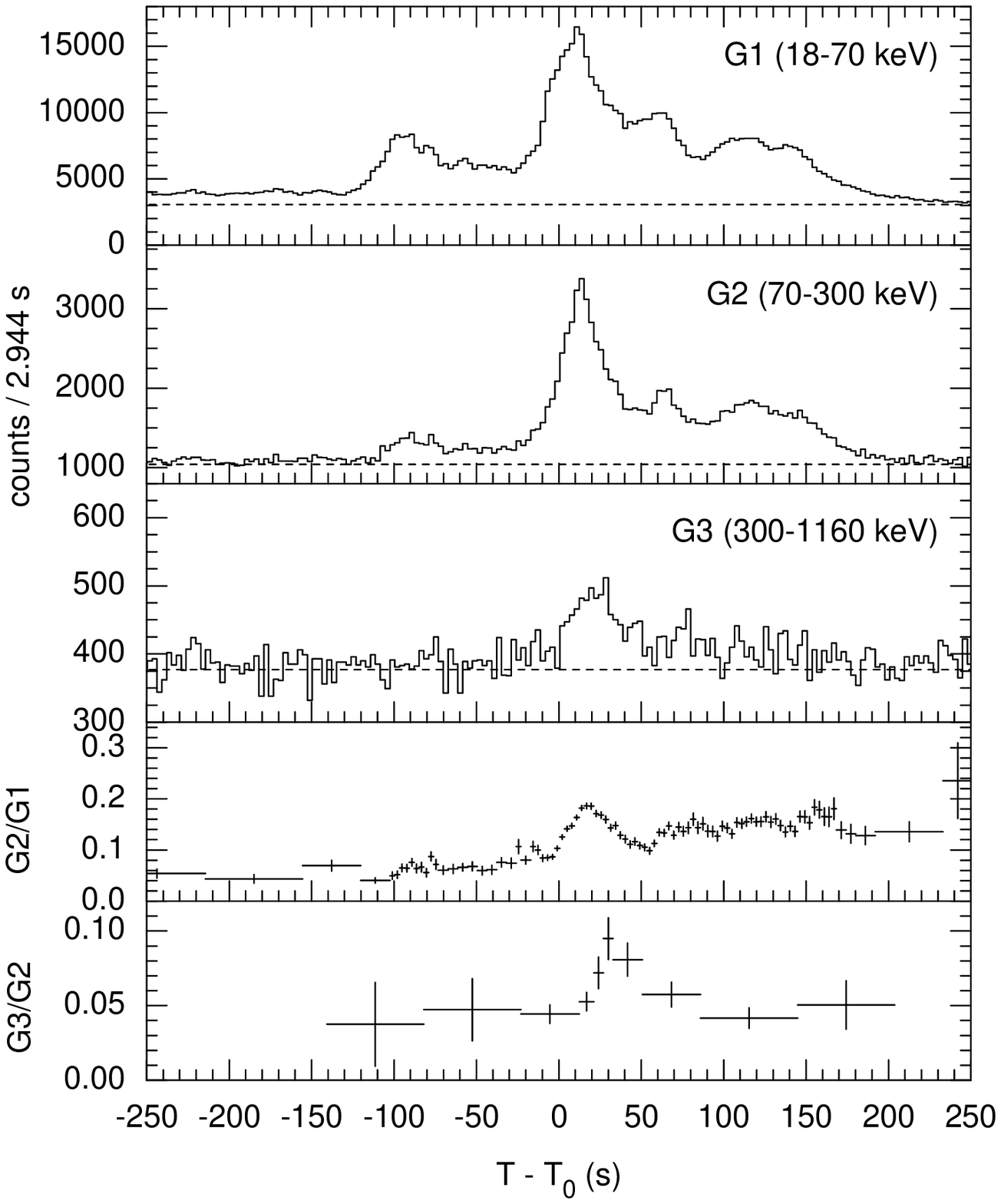}
\hfill
\includegraphics[width=0.47\textwidth]{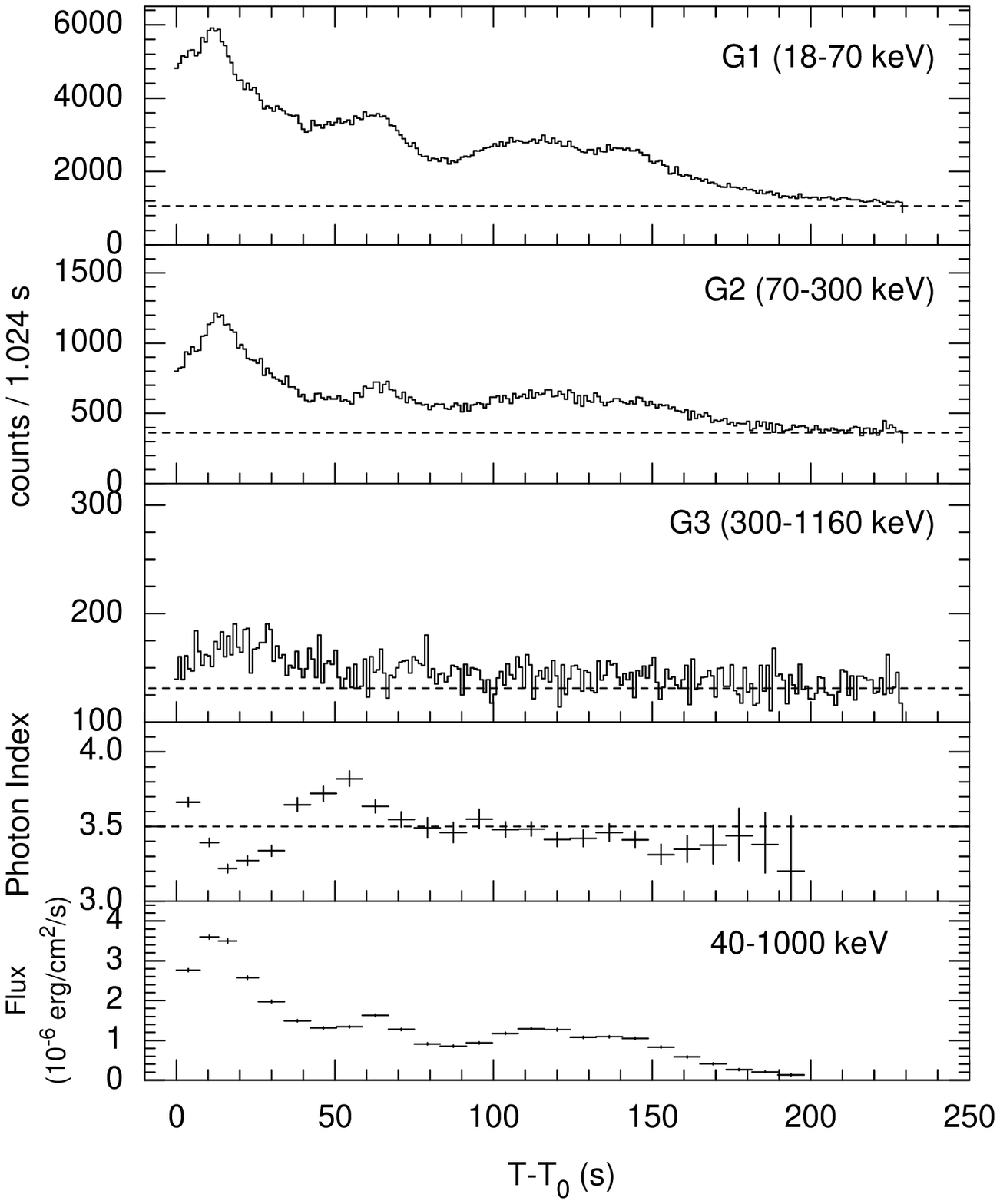}
\caption{Solar Flare on November 8, 2012 (class M1.7).
T$_0$=8365.302 s UT (02:19:25.302).}
\label{Fig3}
\end{figure}

Figure~\ref{Fig3} shows a solar flare recorded by
Konus-\textit{Wind} on November 8, 2012 (class M1.7). The left panel
shows the time profile of the flare in three energy channels and two
hardness ratios: G2/G1 and G3/G2, produced from background data; on
the right -- time profile with a resolution of 1.024~s, the
evolution of the power-law spectral index and the energy flux in the
40--1000~keV range obtained from ``Burst'' mode data. It can be seen
that the flare starts with a fairly soft emission, followed by a
powerful hard pulse (which caused the trigger event) during which a
correlation between intensity and hardness is clearly seen, then the
intensity of the radiation gradually decreases, but its hardness
remains almost constant with the spectral index of $\sim$3.5.

\subsection{Flare SOL2002-03-10T01:34}
\label{SecSOL20020310}
\begin{figure}
\centering
\includegraphics[height=0.5\textheight]{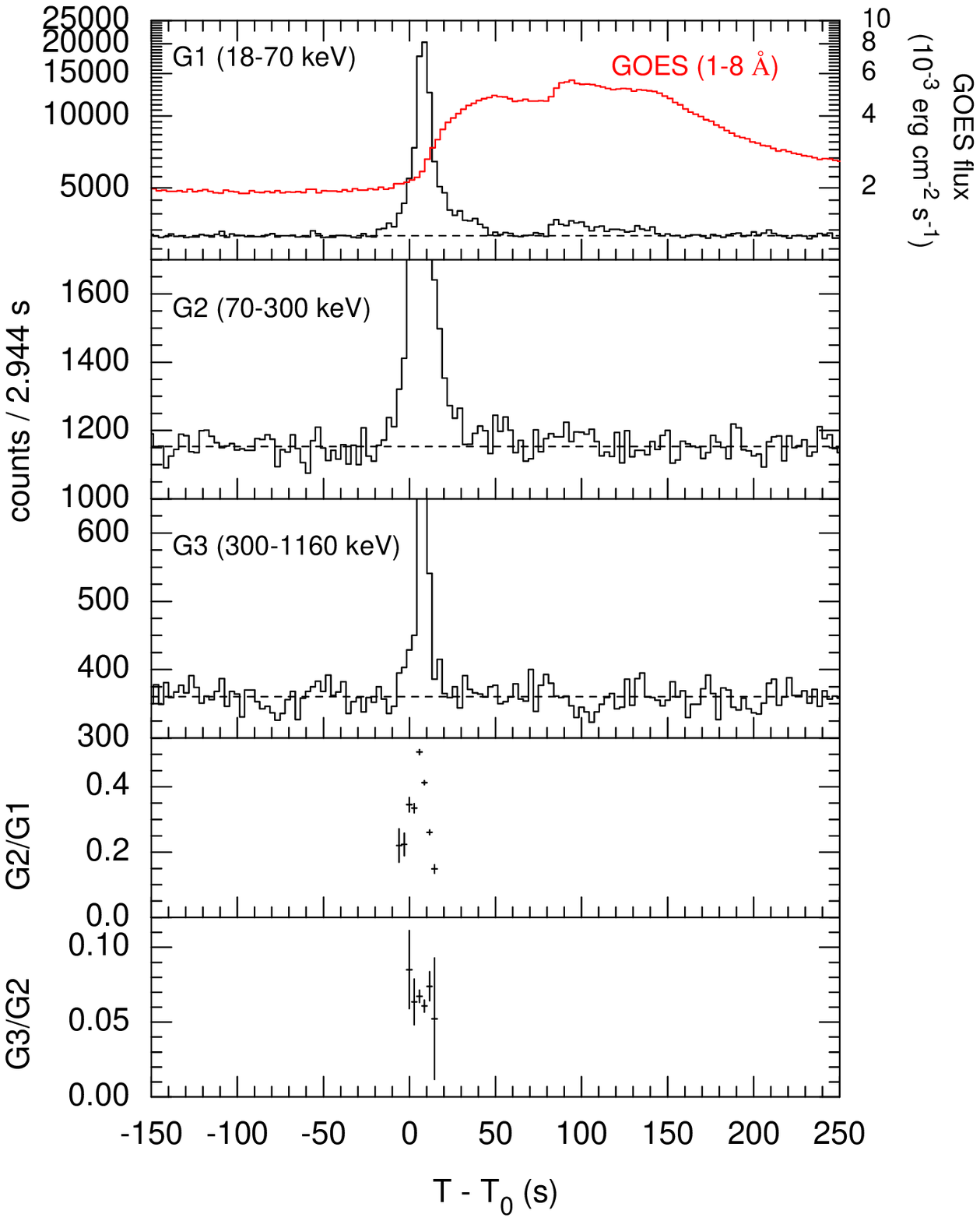}
\hfill
\includegraphics[height=0.5\textheight]{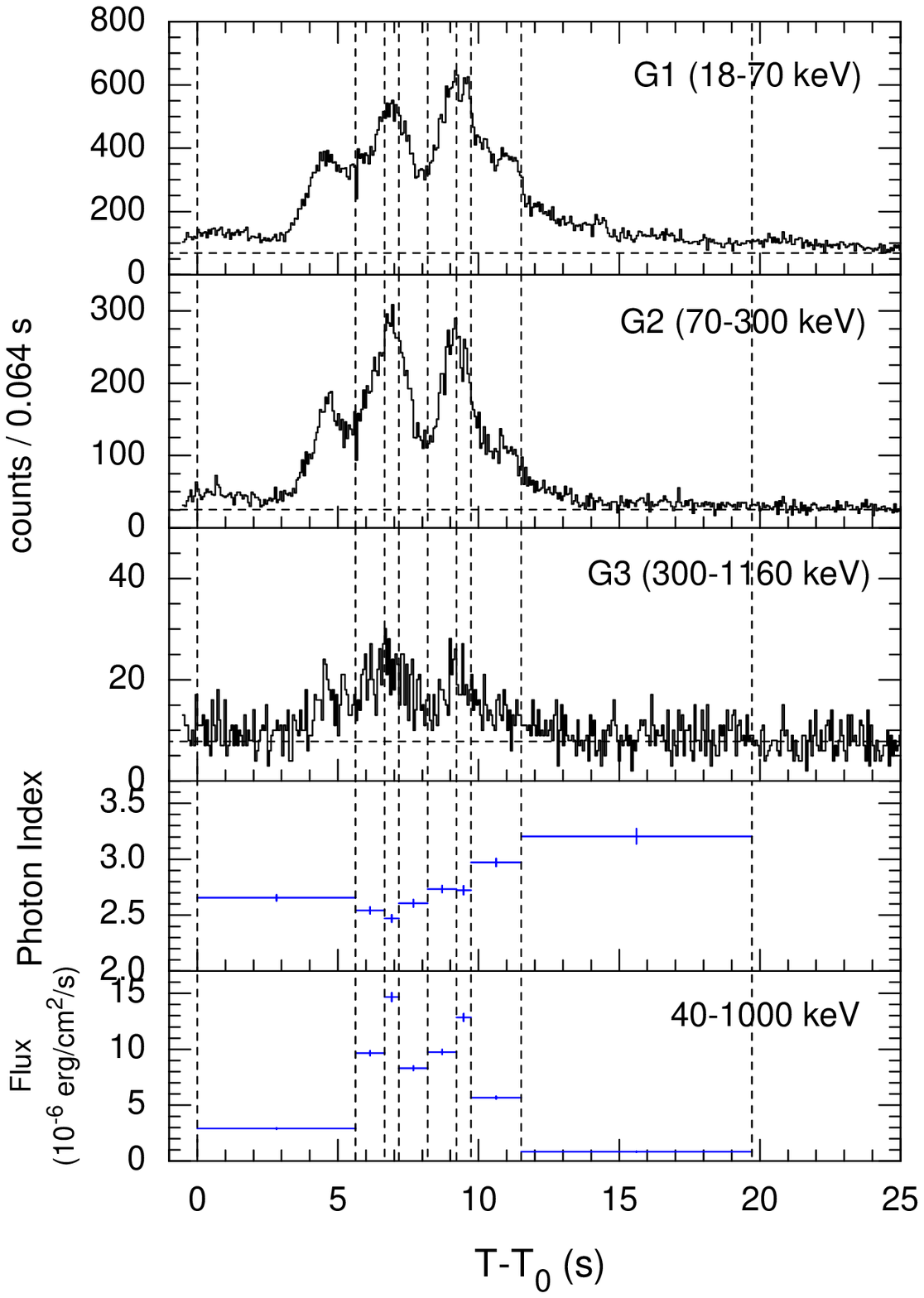}
\caption{Solar Flare on March 10, 2002 (class C5.1) -- ``early
impulsive flare''. T$_0$=5693.874 s UT (01:34:53.874).}
\label{Fig4}
\end{figure}

Figure~\ref{Fig4} (similar to Figure~\ref{Fig3}) shows a solar flare
detected by Konus-\textit{Wind} on March 10, 2002 (class C5.1). The
flare is quite unusual -- the duration of the main pulse is only
about 15~s, without any significant emission before the pulse even
in the soft channel, and the spectrum of the emission is extremely
hard with an index of 2.4 at the maximum intensity. In $\sim$70~s
after the main pulse a much weaker pulse follows that is only
visible in the channel G1 (in this case the pulse has greater
intensity in the GOES channels, as shown in the left panel of
Fig.~\ref{Fig4}). This flare belongs to a class of so-called ``early
impulsive flares'' \citep{Sui2006}. For such flares a hard radiation
peak precedes a soft radiation peak, indicating weak plasma
preheating, while the images obtained by \textit{RHESSI}, typically
exhibit two-footpoint morphology of the source for the hard peak and
looptop source for the delayed soft radiation peak; the spectrum of
the early impulsive peak is nonthermal and very hard and the
spectrum is thermal in the delayed soft radiation pulse (see, for
example, \citet{Su2008}). In this flare the delay between the soft
and hard peaks is $\sim$20~s.

\section{OBSERVATIONS OF SOLAR FLARES IN HELICON-CORONAS-F
EXPERIMENT}

Useful exposure for the Coronas-F observations was severely limited
by passing through the radiation belts at high latitudes and through
the South Atlantic Anomaly (SAA). In total, for 4 years of the
experiment 300 flares had been registered in the trigger mode: 4
X-class, 54 M-class, 163 C-class and 69 B-class (flares were counted
providing their trigger fired on the portions of orbit with a modest
background  level, i.e. at low and moderate latitudes away from the
SAA). The flares were registered by the detector whose axis was
oriented to the Sun, herewith, the radiation passed through an
entrance beryllium window that provided low absorption in the soft
part of the spectrum. Measurements of 256-channel spectra in the
``Background'' mode (see section~\ref{SecHel}) allowed registering
the deuterium line 2.223~MeV and studying its evolution on the
timescale of 33.6~s in strong long flares.

\subsection{Flare SOL2003-10-26T06:11}
\begin{figure}
\centering
\includegraphics[width=0.8\textwidth]{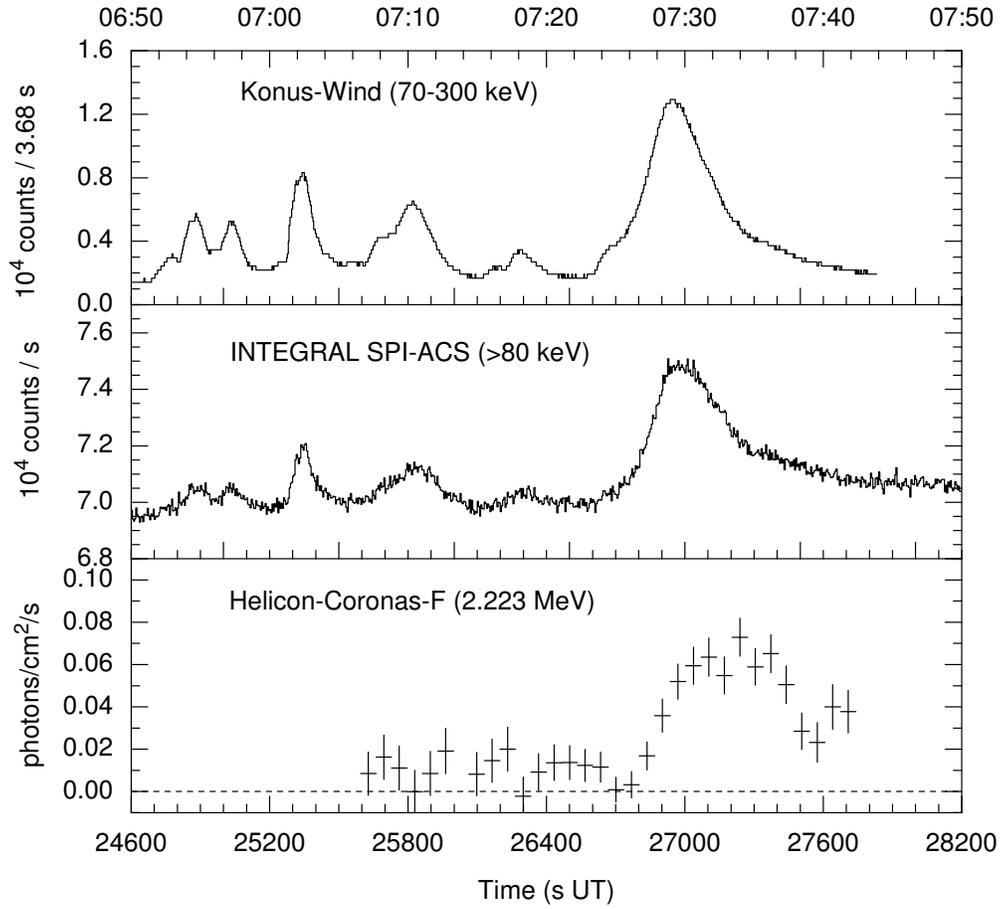}
\caption{Time profile of the solar flare on October 26, 2003 drawn
from the Konus-\textit{Wind} and \textit{INTEGRAL} SPI-ACS data and
the evolution of the 2.223~MeV line intensity derived from the
Helicon data.}
\label{Fig5}
\end{figure}

Figure~\ref{Fig5} shows a solar flare detected on October 26, 2003
(class X1.2). The flare is interesting because after $\sim$40 min
after its start a powerful pulse of hard emission was detected,
which was not accompanied by any increase in the intensity of soft
X-rays \citep{Zimovets2012}. The top panels of Fig.~\ref{Fig5} show
the time profile derived from the data of the Konus-\textit{Wind}
standby system (see section \ref{secKW}) and from the
anti-coincidence shield (ACS) of the SPI telescope onboard the
\textit{INTEGRAL} observatory. The bottom panel shows the emission
intensity evolution in the deuterium line, derived from the Helicon
data. It can be seen that the line intensity is following the
intensity of the powerful hard X-rays pulse (with a peak at
$\sim$07:30 UT) with a delay of about 100~s needed to thermalize
neutrons.

\begin{figure}
\centering
\includegraphics[width=0.8\textwidth]{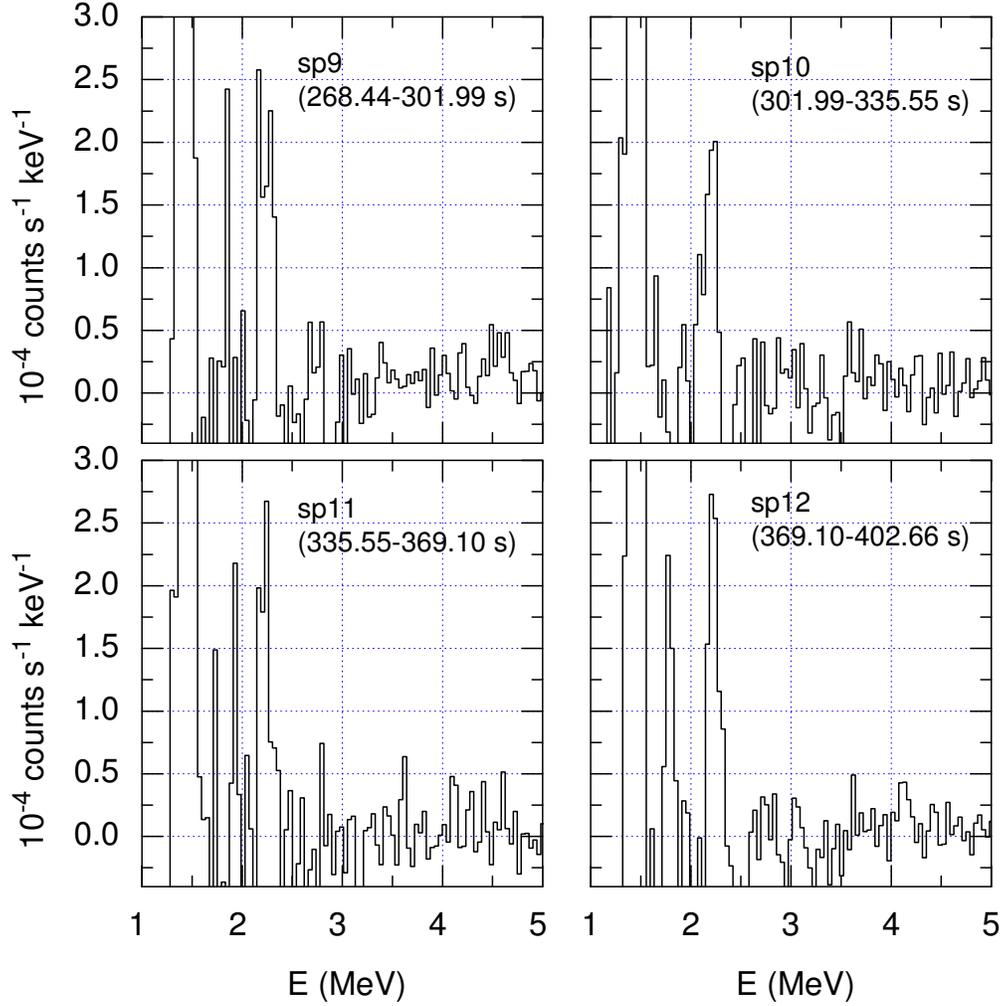}
\caption{Four consecutive spectra measured by Helicon during the
flare SOL2003-10-26T06:11 (continuum subtracted). The spectrum
accumulation times are given relative to the Konus-\textit{Wind}
trigger time T$_0$=22285.091 s UT (06:11:25.091). The spectrum shows
the 2.223~MeV neutron capture line and the strong 1.460~MeV line
caused by the decay of $^{40}$K present in the detector materials.}
\label{Fig6}
\end{figure}

To calculate the intensity in the line, it is not necessary to know
a background spectrum, it is enough to subtract the continuum
defined by the nearby regions of the spectrum that do not contain
lines. Figure~\ref{Fig6} shows the individual spectra (with the
continuum subtracted) measured by Helicon in the 1--5 MeV~range. The
spectra demonstrate the 2.223~MeV line, an escape peak corresponding
to the energy 1.712 MeV (=2.223-0.511), and the strong 1.460 MeV
line caused by the decay of $^{40}$K present in the detector
materials. The deuterium line width is determined by the resolution
of the instrument that amounts to 130 keV at 2.223 MeV ($\simeq
6\%$; FWHM) (wherein the width of the spectral channels in this
range is 40~keV). To convert the line intensity measured in
counts~s$^{-1}$ to the photon flux (photons~cm$^{-2}$~s$^{-1}$), the
measured intensity was divided by the detector effective area in the
full absorption peak, which is equal to 32 cm$^2$ at 2.223 MeV. The
line fluence measured during the hard pulse (906~s since 07:27:16
UT), was 22.1$\pm$0.8~photons~cm$^{-2}$, which is comparable with
the fluences measured with the GRS gamma-ray spectrometer onboard
the Solar Maximum Mission in the 2.223~MeV line for the most intense
solar flares \citep{Vestrand1999}.

\section{CONCLUSIONS}
In the Konus-\textit{Wind} experiment, a large database of solar
flares has been accumulated since 1994 upto the present (834 flares
in the ``Burst'' mode). These data allow us to investigate the
dynamics of hard X-ray and gamma-ray emission spectra in different
phases of flares on timescales varying from 0.256 to 8.2~s (but the
initial rise phase is often skipped), as well as the temporal
evolution and the evolution of the hardness ratio on the timescales
from 2 to 256~ms in the ``Burst'' mode and on the timescale of
2.994~s in the ``Background'' mode.

The Helicon Coronas-F data allow studying the dynamics of the
intensity of the 2.223~MeV deuterium line and its relationship to
the intensity of hard X-ray and gamma radiation.

A list of the Konus-\textit{Wind} solar flare triggers and figures
of their time profiles are available at
http://www.ioffe.ru/LEA/Solar/.

\acknowledgements This work is partially supported by PR22 FTSPK 1.5
N8524.


%

%



\begin{thebibliography}{}
%
\bibitem[Aptekar et al.(1995)]{Aptekar1995} Aptekar, R., Frederiks, D., Golenetskii, S., et al. 1995, \ssr \, 71, 265
%
\bibitem[Arnaud (1996)]{Arnaud1996} Arnaud, K. A. 1996, in ASP Conf. Ser. 101,
Astronomical Data Analysis Software and Systems V, ed. G. Jacoby \&
J. Barnes (San Francisco, CA: ASP), 17
%
\bibitem[Fletcher et al.(2011)]{Fletcher2011} Fletcher, L., Dennis, B.R., Hudson, H.S., et al. 2011, \ssr, 159, 19
%
\bibitem[Lin et al.(2002)]{Lin2002} Lin, R.P., Dennis, B.R., Hurford, G.J., et al. 2002,
\solphys, 210, 3
%
\bibitem[Mazets et al.(2012)]{Mazets2012} Mazets, E.P., Aptekar, R.L., Golenetskii, S.V., et al. 2012, JETP
Letters, 96, 544
%
\bibitem[Meegan et al.(2009)]{Meegan2009} Meegan, C., Lichti, G., Bhat, P., et al. 2009, \apj, 702, 791
%
\bibitem[Oraevskii et al.(2002)]{Oraevskii2002} Oraevskii, V.N., Sobel'man, I.I., Zhitnik, I.A., Kuznetsov, V.D.
 2002, Phys. Usp., 45, 886
%
\bibitem[Su et al.(2008)]{Su2008} Su, Y., Gan, W.Q., \& Li, Y. P. 2008, AdSpR, 41, 988
%
\bibitem[Sui et al.(2006)]{Sui2006} Sui, L., Holman, G.D., \& Dennis, B.R. 2006, \apjl, 645, L157
%
\bibitem[Vestrand et al.(1999)]{Vestrand1999} Vestrand, W.T., Share, G.H., Murphy, R.J., et al. 1999,
\apjs, 180, 409
%
\bibitem[Zimovets \& Struminsky (2012)]{Zimovets2012} Zimovets, I., \& Struminsky, A. 2012, \solphys, 281, 749
%
\end{thebibliography}
\end{document}